\documentclass[numberedappendix,useAMS,usenatbib,letterpaper]{mn2e}
\usepackage[pass]{geometry}
\usepackage{amssymb,amsmath}
\usepackage{verbatim}
\usepackage{color}
\usepackage{hyperref}
\usepackage{needspace}
\usepackage{enumitem}
\usepackage{tikz}
\usepackage{graphicx}
\usepackage[draft]{minted}
\usetikzlibrary{positioning}
\definecolor{linkcolor}{rgb}{0,0,0.25}
\hypersetup{
  colorlinks=true,        
  linkcolor=linkcolor,    
  citecolor=linkcolor,    
  filecolor=linkcolor,    
  urlcolor=linkcolor      
}
\newcounter{address}
\title[Winding transient spirals in \emph{Gaia} DR2]
{Transient spiral structure and the disc velocity substructure in \emph{Gaia} DR2}

\author[J. A. S. Hunt et al.]
{\parbox{\textwidth}{Jason A.~S.~Hunt$^1$, Jack Hong$^{2,3}$, Jo~Bovy$^{1,3,4}$, Daisuke Kawata$^{5}$ and Robert J. J. Grand$^{6,7}$}\vspace{0.5cm}
\\
$^{1}$ Dunlap Institute for Astronomy and Astrophysics, University of Toronto, 50 St. George Street, Toronto, Ontario, M5S 3H4, Canada\\
$^{2}$ Department of Physics and Astronomy, University of British Columbia, 6224 Agricultural Road, Vancouver, BC V6T 1Z1, Canada\\
$^{3}$ Department of Astronomy and Astrophysics, University of Toronto, 50 St. George Street, Toronto, ON, M5S 3H4, Canada \\
$^{4}$ Alfred P. Sloan Fellow\\
$^{5}$ Mullard Space Science Laboratory, University College London, Holmbury St. Mary, Dorking, RH5 6NT, UK\\
$^{6}$ Heidelberger Institut f\"ur Theoretische Studien, Schloss-Wolfsbrunnenweg 35, 69118 Heidelberg, Germany\\
$^7$ Zentrum f\"{u}r Astronomie der Universit\"at Heidelberg, Astronomisches Recheninstitut, M\"onchhofstr. 12-14, 69120 Heidelberg, Germany
}

\pagerange{\pageref{firstpage}--\pageref{lastpage}}
\pubyear{2017}

\begin{document}

\maketitle

\label{firstpage}

\begin{abstract}
The second data release from ESA's $Gaia$ mission has revealed many ridge-like structures in the velocity distribution of the Milky Way. We show that these can arise naturally from winding transient spiral structure that is commonly seen in $N$-body simulations of disk galaxies. We construct test particle models of the winding spiral structure, and compare the resulting distribution of orbits with the observed two-dimensional velocity distribution in the extended solar neighbourhood and with the distribution of rotational velocities over 8 kpc along the Sun--Galactic-centre--Galactic anti-centre line. We show that the ridges in these observations are well reproduced by the winding spiral model. Additionally, we demonstrate that the transient winding spiral potential can create a Hercules-like feature in the kinematics of the solar neighbourhood, either alone, or in combination with a long-slow bar potential.
\end{abstract}


\section{Introduction}\label{intro}
The recent second data release \citep[DR2;][]{DR2} from the European Space Agency's $Gaia$ mission \citep{GaiaMission} provides a new window on the dynamics of the Solar neighbourhood. DR2 contains $\sim1.6\times10^9$ stars, with 5 parameter phase space information for $\sim1.3\times10^9$ of those stars, and 6 parameter phase space information (i.e. including radial velocities) for $\sim7\times10^6$ of those stars. 

With this new wealth of information we are able to trace the kinematics of the disc across multiple kpc \citep[e.g.][]{GCKatz+18}. One of the more striking discoveries is the presence of ripples in the velocity distribution, e.g. \cite{KBCCGHS18,RAF18} show the `ridges' present in the distribution of Galactocentric rotation velocities $v_{\phi}$ with radius, and \cite{Antoja+18} show the presence of `arches' and `shells' in the $U$-$V$-$W$ planes, and `spiral' features in the $Z$-$W$ distribution. This clearly indicates that the Milky Way disc is not in equilibrium, and has been recently perturbed \citep[although note that vertical perturbations have been observed previously, e.g.][so this is not entirely unexpected]{WGYDC12}. Potential explanations include the previous passage of the Sagittarius dwarf galaxy, or a recent merger \citep[e.g. as suggested by][]{MQWFNSB09}, or the impact of the Galactic spiral structure. For example, \cite{Quillen+18} propose that the ridges trace orbits for stars which have recently crossed nearby spiral arms, linking the ridges from specific arms with the divisions between the moving groups in the Solar neighbourhood. However, these ridges also look similar to the structure predicted in the models of \cite{dSWT04}, which contain transient spiral waves. While they do not reproduce the curvature of the arches observed in e.g. \cite{Antoja+18,RAF18}, they are qualitatively similar in nature and may offer an explanation.

It is known that the spiral structure has a significant effect on the kinematics in the Solar neighbourhood. For example, \cite{QM05} show that a two armed spiral density wave with its 4:1 Inner Lindblad Resonance (ILR) near the Sun can lead to closed orbits which give rise to the Hyades/Pleiades and Coma Berenices moving groups. Similarly, \cite{Sellwood2010} find that stars in the Hyades stream have both action and angle variables in keeping with their having been scattered by the ILR of a spiral potential. \cite{MVB17} perform a more generalised exploration of the resonances arising from spiral structure, quantifying the areas of resonant trapping, and chaos across the disc. When modelling the bar and spiral together, their resonances also overlap leading to additional effects. For example \cite{Quillen03} perform a detailed examination of kinematics arising from the resonant overlap of a short fast bar and a spiral pattern, finding both areas of chaos and areas of resonant trapping which could potentially explain Hercules or other streams. 

Most previous works investigating the Hercules stream as a resonant feature of the bar \citep[e.g.][]{Aetal14-2,Monari+16,Hunt+18}, or  the interaction of bar and spiral potentials \citep[e.g.][]{C07,AVPMFF09,MFSGKB16} have focused on a short fast bar model. However, some recent measurements of the bar length favour a longer bar \citep[e.g.][]{WGP15}, which in turn must be a slower bar, for it may not extend past corotation \citep[e.g.][]{Contopoulos80}. 

\cite{QDBMC11} explore the disc kinematics arising from the interaction of a long bar and spiral structure in an $N$-body hybrid simulation. They find `Hercules' around 10 kpc, corresponding to the bar's OLR, but also a number of kinematic features across the disc originating from the spiral structure, which highlights their importance when considering the origin of the streams. \cite{P-VPWG17} constructed an $N$-body model of the Milky Way using the Made-to-Measure method \citep{ST96}, and showed that a long slow bar can reproduce the Hercules stream if stars orbiting the bar's Lagrange points, $L_4$ \& $L_5$ move outwards from corotation and reach the Solar neighbourhood. In \cite{HB18} we used the test particle backwards integration technique from \cite{D00} to show that a long bar with an $m=4$ component could create a Hercules-like stream in the Solar neighbourhood through the 4:1 Outer Lindblad Resonance (OLR), and \cite{Hattori+18} used test particle simulations to show that the combination of bar and spiral structure is able to reproduce Hercules for both a long and short bar. Thus, to truly understand which resonance gives rise to the Hercules stream---or any other resonance observed in the extended solar neighbourhood---requires tracing the stream's location in the velocity distribution over $\gtrsim1$ radian in azimuth in the plane of the disk \citep[e.g.][]{Bovy10,Hunt+18,HB18}

The nature of the spiral structure itself remains the subject of debate. For example, it is known that stars in the inner region of disc galaxies rotate faster than those in the outer regions, and thus, if spiral arms rotate at the same speed as their constituent stars they should wind up over time and be disrupted. This is contrary to many observations of `grand design' spirals in external disc galaxies, and is known as the winding dilemma \citep[e.g.][]{W96}. \cite{LS64} proposed a solution to this winding dilemma by suggesting that spiral arms rigidly rotate through a galactic disc independently of the stars as a long-lived spiral density wave; the spiral pattern speed is assumed to be constant as a function of radius. 

However, $N$-body galaxy models are unable to reproduce this classical density wave-like behaviour despite significant increases in computational power and resolution \citep[e.g.][]{S11,DB14}. Thus, in recent years the transient reforming arms seen in $N$-body simulations have been revisited as a likely explanation for the origin of the spiral structure. They can be explained as the superposition or coupling of long lived spiral modes \citep[as shown in, e.g.,][]{QDBMC11,CQ12,SC14}, which individually behave as a wave of constant pattern speed, but collectively produce transient density enhancements at radii where they overlap, with pattern speeds intermediate to those of the modes by which they are bound. Alternatively, they can be interpreted as a fully corotating material arm \citep[e.g.][]{WBS11,GKC12,GKC12-2}. The non-linear growth of a corotating spiral arm, via a mechanism similar to swing amplification, is difficult to explain via the superposition of spiral modes \citep{GKC12,Kumamoto+Noguchi16}.

In \cite{KHGPC14}, we explored the kinematics on either side of a transient, corotating $N$-body spiral arm, and observed ridges similar to what is seen in $Gaia$ DR2. In \cite{HKGMPC15}, we showed that these features would be visible in the $Gaia$ data, and in \cite{HKMGFS17}, we made an initial detection of the high angular momentum disc stars which form part of the ridge in data from the Tycho-$Gaia$ Astrometric Solution \citep[TGAS;][]{Michalik+15} from $Gaia$ DR1 \citep{GaiaDR1}.

In this work, we investigate the impact of a transient corotating spiral arm potential on the kinematics of the extended Solar neighbourhood, and compare the models with data from $Gaia$ DR2. We show that the transient and winding spiral arms are able to fit the `ridges' or `ripples' in the $R$-$v_{\phi}$ distribution, without needing to invoke an external perturbation, such as from the passage of the Sagittarius dwarf galaxy \citep[e.g. as suggested in][, and also many works on the vertical structure of the disc]{Antoja+18}.

We also find that the winding spiral potential enables us to produce a fit to Hercules either alone, or when combined the long slow bar potential. As previously shown in \cite{Hattori+18}, we conclude that it is difficult to infer the length and pattern speed of the bar from the Hercules stream alone. We stress here that we are not attempting to present the correct parameterisation of the Milky Way's spiral structure, or fit the features exactly. We merely show that a transient winding arm naturally reproduces arches and ridges in the velocity distribution, as observed in $Gaia$ DR2.

\begin{figure}
\centering
\includegraphics[width=0.9\hsize]{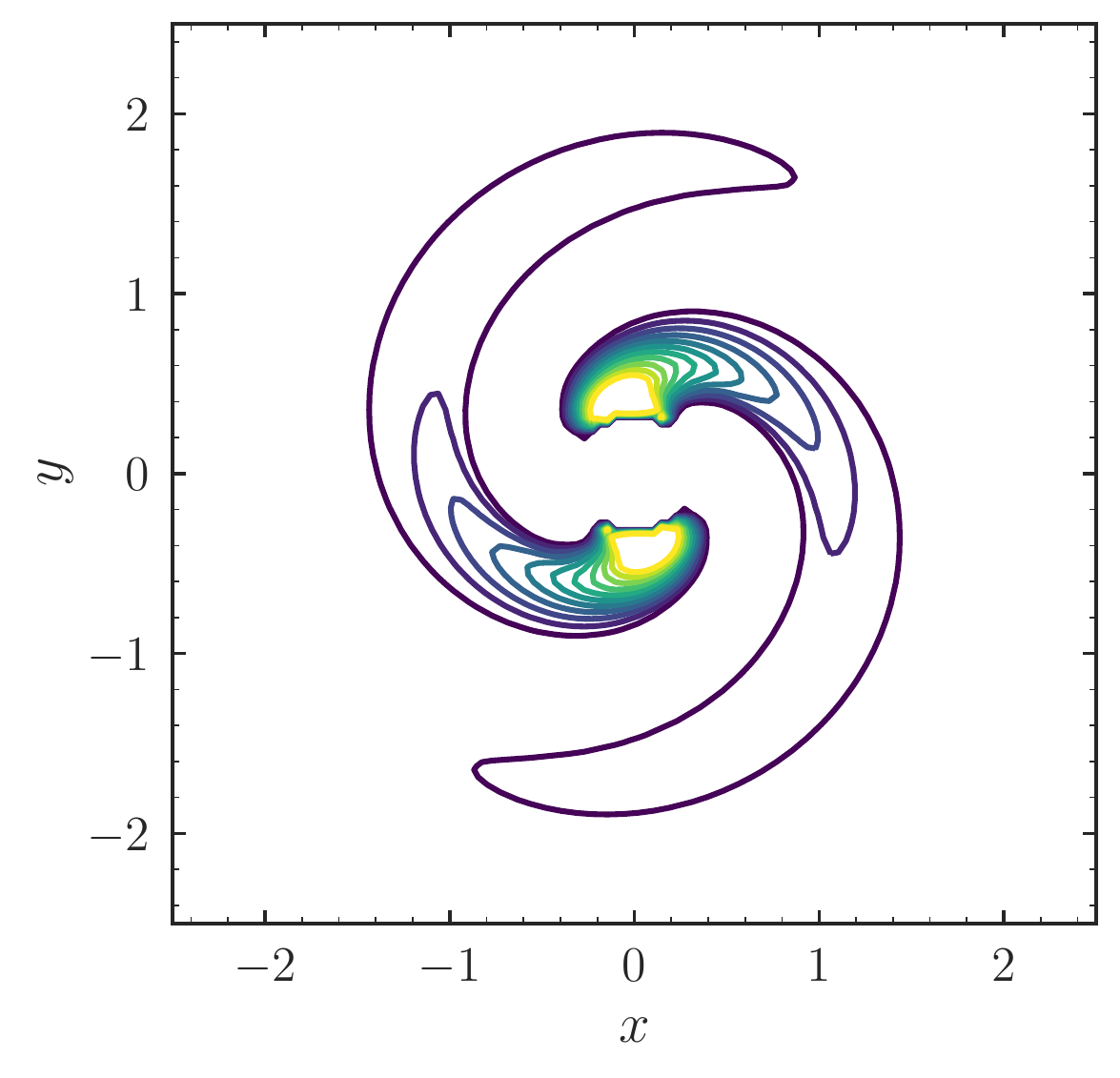}
\includegraphics[width=0.9\hsize]{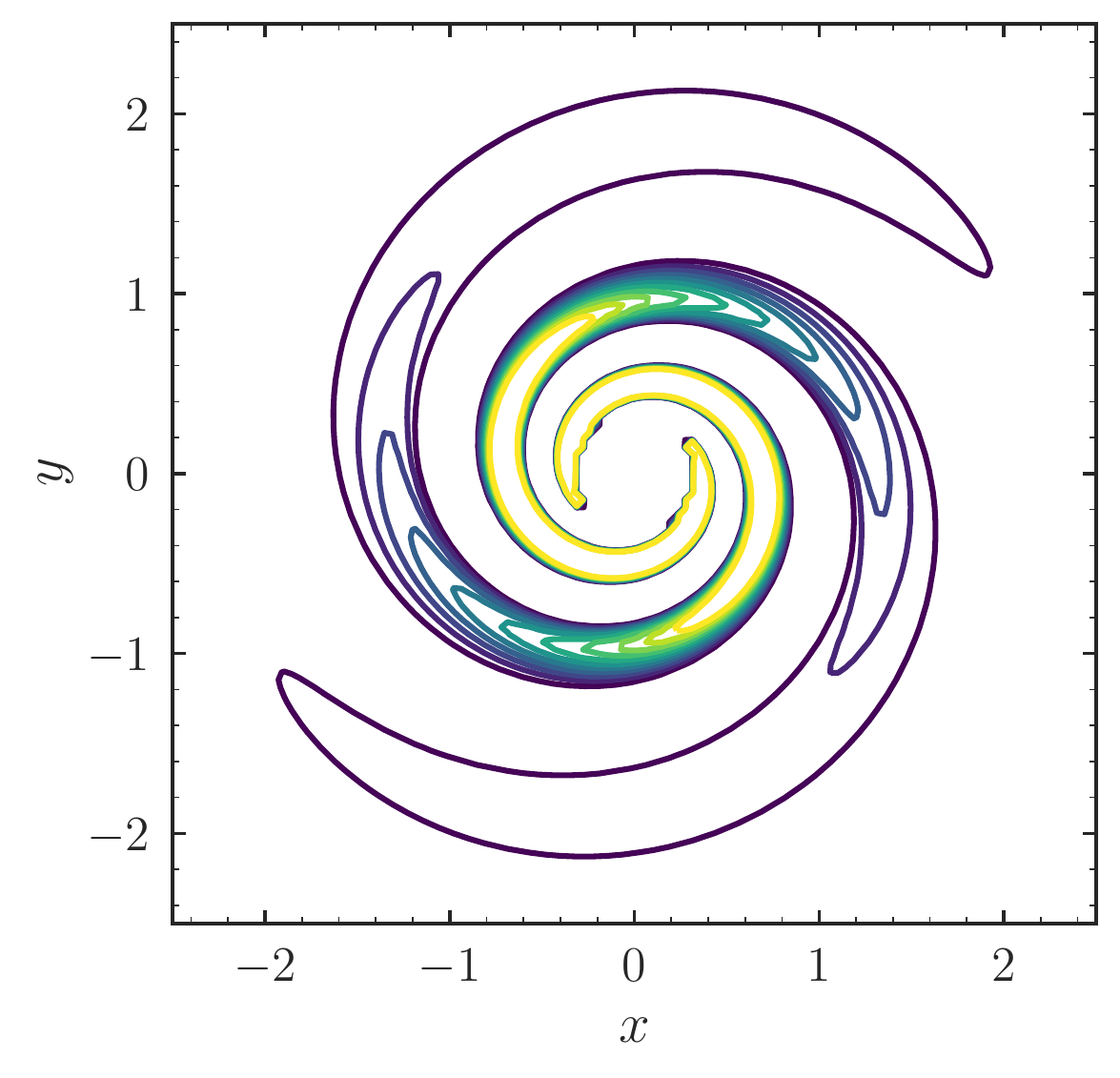}
\includegraphics[width=0.9\hsize]{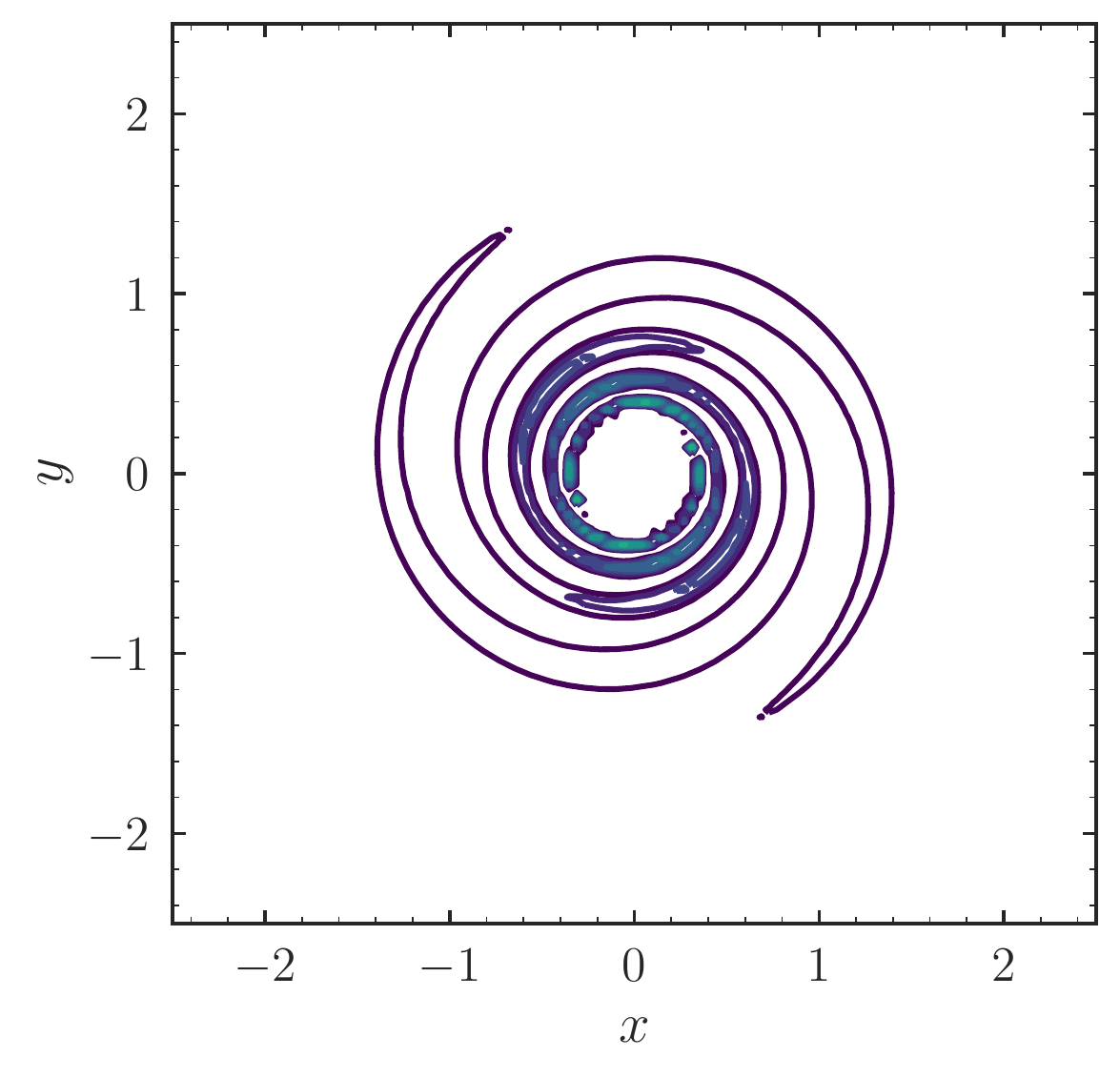}
\caption{Example Density enhancement from the \texttt{SpiralArmsPotential} combined with the \texttt{CorotatingRotationWrapperPotential} for a spiral in the growth phase (upper), the mid point of its lifetime (centre) and the disruption phase (lower), in simulation units such that the distance to the Galactic centre, $R_0=1$.}
\label{threespirals}
\end{figure}

In Section \ref{potential} we describe the disc dynamical model used, including the introduction of both the density-wave and corotating spiral arm potential. In Section \ref{thedata} we demonstrate the models ability to fit multiple moving groups and ripples in the velocity distribution, and in Section \ref{summary} we summarize our results.

\section{The disc dynamics model}\label{potential}
\subsection{Basic setup and bar potential}
To make predictions of the velocity distribution in the Solar neighbourhood, resulting from resonant interaction with the Galactic bar and spiral structure, we use \texttt{galpy}\footnote{Available at \url{https://github.com/jobovy/galpy}~.} \citep{B15} to simulate the distribution of stellar orbits in the disk and the effect of non-axisymmetry on this distribution. In this paper, we only compare to motions in the midplane of the Galaxy and therefore build a two-dimensional model of the orbits in the Milky Way disk.

As in \cite{Hunt+18} and \cite{HB18} we use a Dehnen distribution function \citep{Dehnen99}, which is a function of energy, $E$, and angular momentum, $L$, to model the stellar disc before bar and spiral formation, and represent the distribution of stellar orbits such that
\begin{equation}
f_{\text{dehnen}}(E,L)\ \propto \frac{\Sigma(R_e)}{\sigma^2_{\text{R}}(R_e)}\exp\biggl[\frac{\Omega(R_e)[L-L_c(E)]}{\sigma^2_{\text{R}}(R_e)}\biggr],
\label{DDF}
\end{equation}
where $L_c$, $\Omega(R_e)$ and $R_e$, are the angular momentum, angular frequency and radius, respectively, of a circular orbit with energy $E$. The gravitational potential is assumed to be a simple power-law, such that the circular velocity is given by
\begin{equation}
  v_c(R)=v_0(R/R_0)^{\beta}\,,
\end{equation}
where $v_0$ is the circular velocity at the solar circle at radius $R_0$.

To model the bar in all models we use the general form of the $\cos(m\phi)$ potential shown in \cite{HB18}, adapted from the quadrupole potential from \citet{D00} and repeated here for convenience; in \texttt{galpy}, this is the \texttt{CosmphiDiskPotential} model. The bar potential is given by
\begin{equation}
\begin{split}
&\Phi_{\mathrm{b}}(R,\phi)=A_{\text{b}}(t)\cos(m(\phi-\phi_{\text{b}}t))\\ 
& \quad \quad \times
\left\{ \begin{array}{ll} -(R/R_0)^p, & \mathrm{for}\ R \geq R_{\text{b}},\\ ([R_{\text{b}}/R]^p-2)\times(R_{\mathrm{b}}/R_0)^p, & \mathrm{for}\ R \leq R_{\text{b}}, \end{array}
\right.
\end{split}
\end{equation}
where the bar radius, $R_{\text{b}}$, is set to $80\%$ of the corotation radius, and $\phi_{\mathrm{b}}$ is the angle of the bar with respect to the Sun--Galactic-center line. The potential is equivalent to the \cite{D00} quadrupole bar for $m=2$ and $p=-3$, where $m$ is the integer multiple of the $\cos$ term, and $p$ is the power law index.

\begin{figure*}
\centering
\includegraphics[width=\hsize]{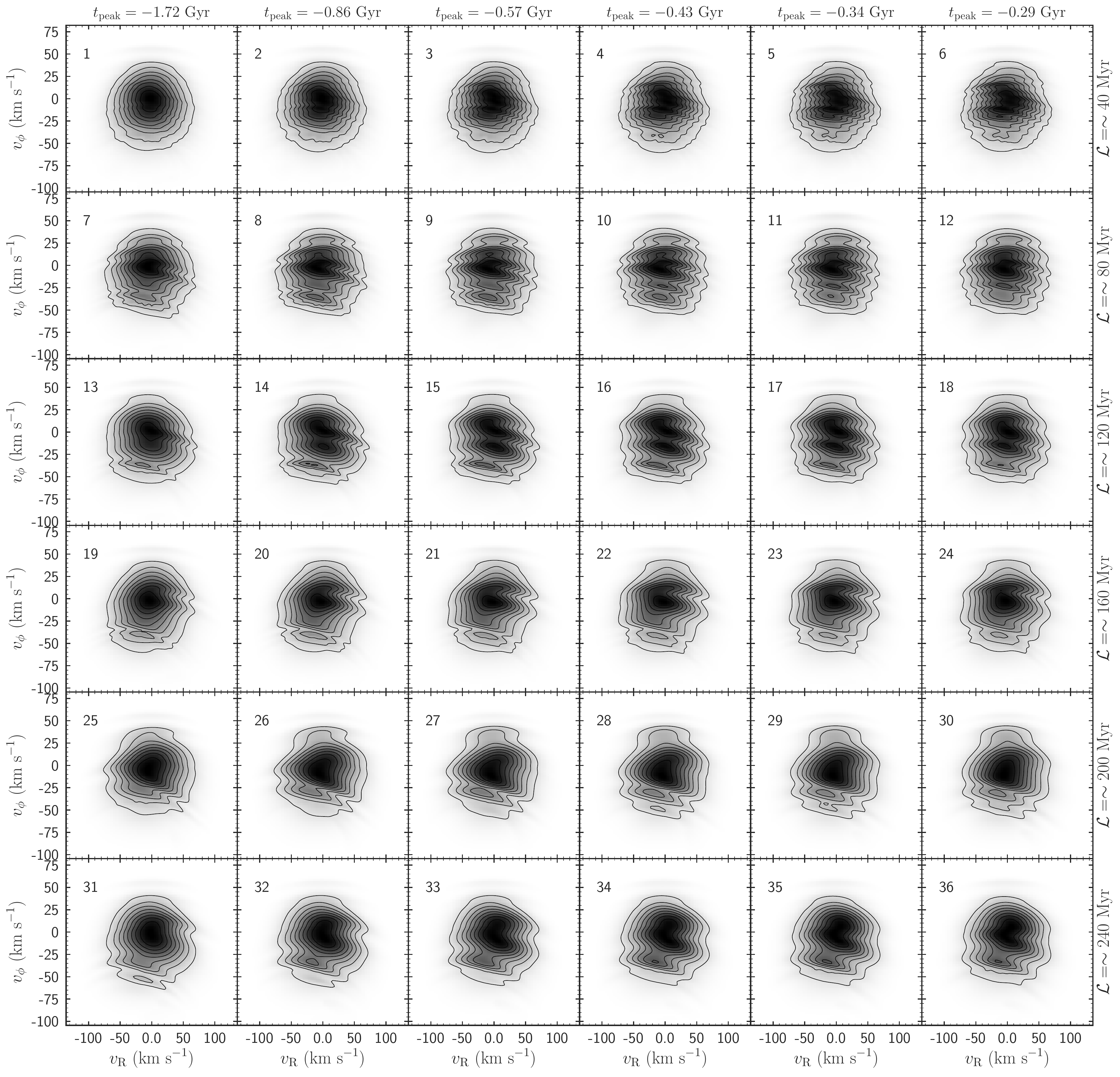}
\caption{$v_{\text{R}}$-$v_{\phi}$ plane for 36 models with increasing spiral arm lifetime, $\mathcal{L}$ (top to bottom), and decreasing time since the peak of the spiral arm density enhancement, $t_{\mathrm{peak}}$ (left to right), such that the formation time, $t_{\mathrm{form}}=t_{\mathrm{peak}}-\mathcal{L}/2$, for a spiral arm potential in combination with a long bar potential.}
\label{LvsF}
\end{figure*}

We grow the bar smoothly such that
\begin{eqnarray}
A_{\text{b}}(t)=
\left\{\begin{array}{ll} 0,\ \frac{t}{T_{\text{b}}}<t_{\text{1}} \\ A_f\biggl[\frac{3}{16}\xi^5-\frac{5}{8}\xi^3+\frac{15}{16}\xi+\frac{1}{2}\biggr], t_{\text{1}}\leq\frac{t}{T_{\text{b}}}\leq t_{\text{1}}+t_{\text{2}}, \\ A_f,\ \frac{t}{T_{\text{b}}} > t_{\text{1}} + t_{\text{2}}.  \end{array}
\right.\,
\end{eqnarray}
where $t_1$ is the start of bar growth, set to half the integration time, $t_2$ is the duration of the bar growth and $T_{\text{b}}=2\pi/\Omega_{\text{b}}$ is the bar period such that
\begin{equation}
\xi=2\frac{t/T_{\text{b}}-t_{\text{1}}}{t_{\text{2}}}-1,
\end{equation}
and
\begin{equation}
A_f=\alpha_{m}\frac{v_0^2}{3}\biggl(\frac{R_0}{R_b}\biggr)^{3},
\end{equation}
where $\alpha_{m}$ is the dimensionless ratio of forces owing to the $\cos(m\phi)$ component of the bar potential and the axisymmetric background potential, $\Phi_0$, at Galactocentric radius $R_0$ along the bar's major axis. This growth mechanism ensures that the bar amplitude along with it's first and second derivatives are continuous for all $t$, allowing a smooth transition from the non-barred to barred state \citep{D00}.

For the model presented in this paper, we set $R_{\mathrm{b}}=5$ kpc, $\Omega_{\mathrm{b}}=1.3$ km s$^{-1}$ kpc$^{-1}$, $\phi_{\mathrm{b}}=25$ deg and $\alpha_{m=2}=0.01$. The bar strength of $\alpha_{m=2}=0.01$ \citep[following e.g.][]{D00,MFSGKB16} corresponds to the radial force from the bar equaling 1\% of the axisymmetric force. This is on the weaker end of estimates of the Milky Way bar strength, e.g. \cite{2015ApJ...800...83B} found $\alpha\approx1.5\%$ by fitting the power spectrum of velocity fluctuations in the Milky Way disc.

\subsection{Spiral potential}\label{spiral}
As discussed in e.g. \cite{HB18}, the bar-only models, regardless of length and pattern speed, and regardless of their ability to reproduce the Hercules stream, do not well reproduce the kinematic substructure and moving groups in the main mode (e.g. the area excluding Hercules) in the Solar neighbourhood $v_{\text{R}}$-$v_{\phi}$ plane. This is unsurprising because a rigidly-rotating bar only induces a small number of resonance regions in the disc and other non-axisymmetric structures such as the spiral arms are thought to have a significant effect on local kinematics \citep[e.g.][]{QM05,Sellwood2010,MVB17,Hattori+18}. Additionally, the coupling between bar and spiral resonances will likely play an important role in shaping kinematic structure across the Galactic disc \citep[e.g.][]{Quillen03,MFSGKB16}. 

\begin{figure}
\centering
\includegraphics[width=\hsize]{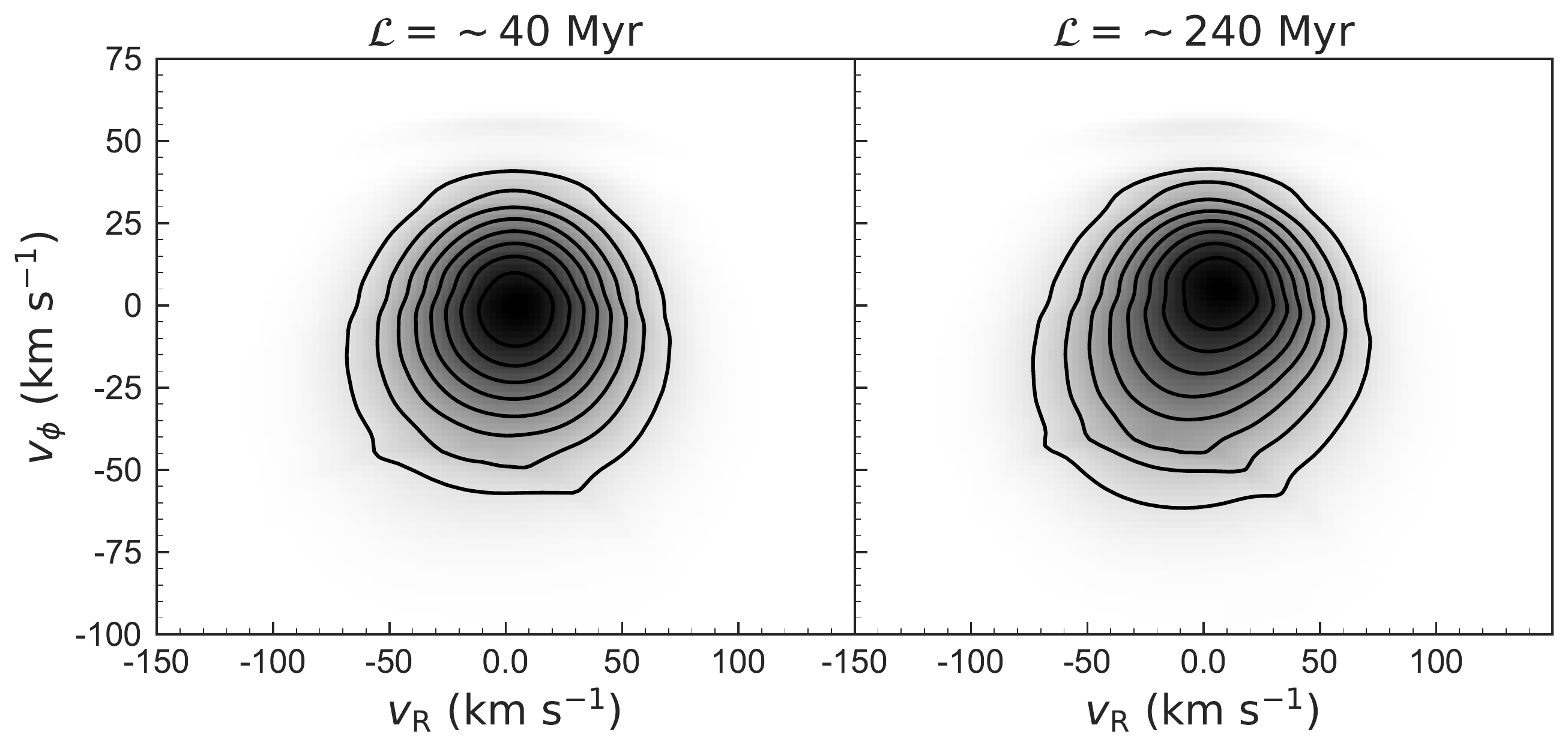}
\caption{$v_{\text{R}}$-$v_{\phi}$ plane in the Solar neighbourhood for models where the peak of the density enhancement occurs at the present day, $t=0$, with lifetime 40 Myr (left) and 240 Myr (right).}
\label{presentday}
\end{figure}

For our spiral arm potential we use the \texttt{SpiralArmsPotential} from \texttt{galpy}, which is an implementation of the sinusoidal potential from \cite{CG02} such that
\begin{eqnarray}
\label{SAP}
\Phi(R,\phi,z)&=&-4\pi GH\rho_0\exp\biggl(\frac{r_{\mathrm{0}}-R}{R_s}\biggr)  \nonumber \\
&\times&\sum\frac{C_n}{K_n D_n}\cos(n\gamma)\biggl[\mathrm{sech}\biggl(\frac{K_n z}{\beta_n}\biggr)\biggr]^{B_n},
\end{eqnarray}
where
\begin{equation}
K_n=\frac{nN}{R\sin(\theta_{\mathrm{sp}})},
\end{equation}
\begin{equation}
B_n=K_nH(1+0.4K_nH),
\end{equation}
\begin{equation}
D_n=\frac{1+K_nH+0.3(K_nH)^2}{1+0.3K_nH},
\end{equation}
\begin{equation}
\gamma=N\biggl[\phi-\phi_{\mathrm{ref}}-\frac{\ln(R/r_{\mathrm{0}})}{\tan(\theta_{\mathrm{sp}})}\biggr],
\end{equation}
$N$ is the number of spiral arms, $\theta_{\mathrm{sp}}$ is the pitch angle, $\rho_0$ is the density at $r_0$, $\phi_{\mathrm{ref}}$ is the reference angle, $R_s$ is the radial scale length of the arm and $H$ is the scale height of the arm. Setting $C_n$ to 1 gives a purely sinusoidal potential profile. Alternatively, setting $C_n=[8/3\pi,1/2,8/15\pi]$ results in a potential which behaves approximately as a cosine squared in the arms, and is flat in the inter-arm region \citep{CG02}. Note that while Equation (\ref{SAP}) gives the full form available in \texttt{galpy}, we use the planar form $\Phi(R,\phi,z=0)$, which sets the sech term to 1.

To make this spiral model into a corotating, winding spiral potential, we wrap the \texttt{SpiralArmsPotential} from Equation (\ref{SAP}) in \texttt{galpy}'s \texttt{CorotatingRotationWrapperPotential}, such that
\begin{equation}
\phi \rightarrow \phi + \frac{V_p(R)}{R} \times \left(t-t_0\right) + a_{\mathrm{p}}
\end{equation}
and
\begin{equation}
V_p(R) = V_{p,0}\,\left(\frac{R}{R_0}\right)^\beta\,,
\end{equation}
where $V_p(R)$ is the circular velocity curve, $t_0$ is the time when the potential is unchanged by the wrapper and $a_{\mathrm{p}}$ is the position angle at time $t_0$. This causes the arm to wind up over time, as seen in $N$-body simulations. This model is designed to mimic the material arms which corotate with the stars at all radii, e.g. as described in \cite{GKC12}.

\begin{figure}
\centering
\includegraphics[width=\hsize]{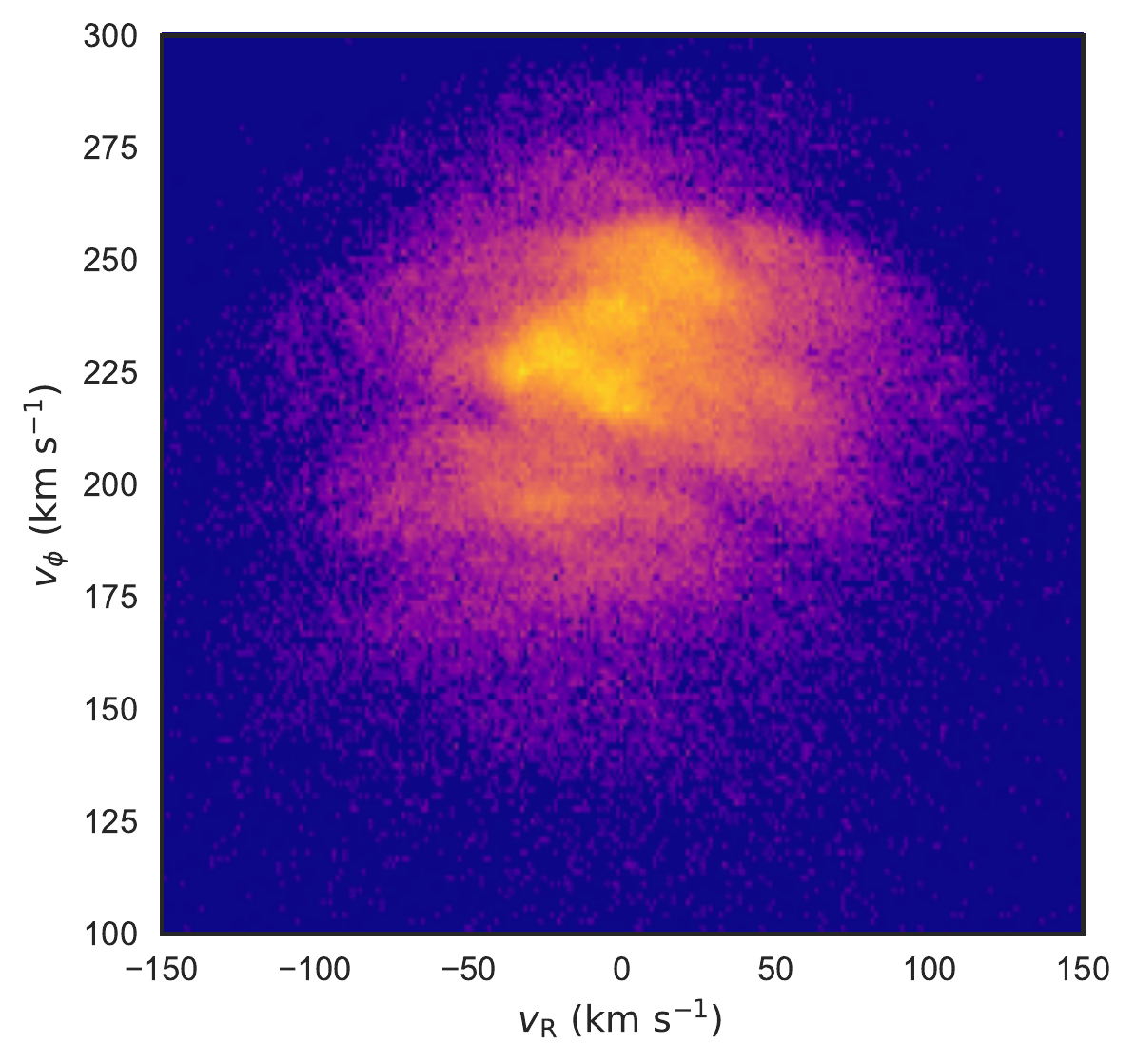}
\includegraphics[width=\hsize]{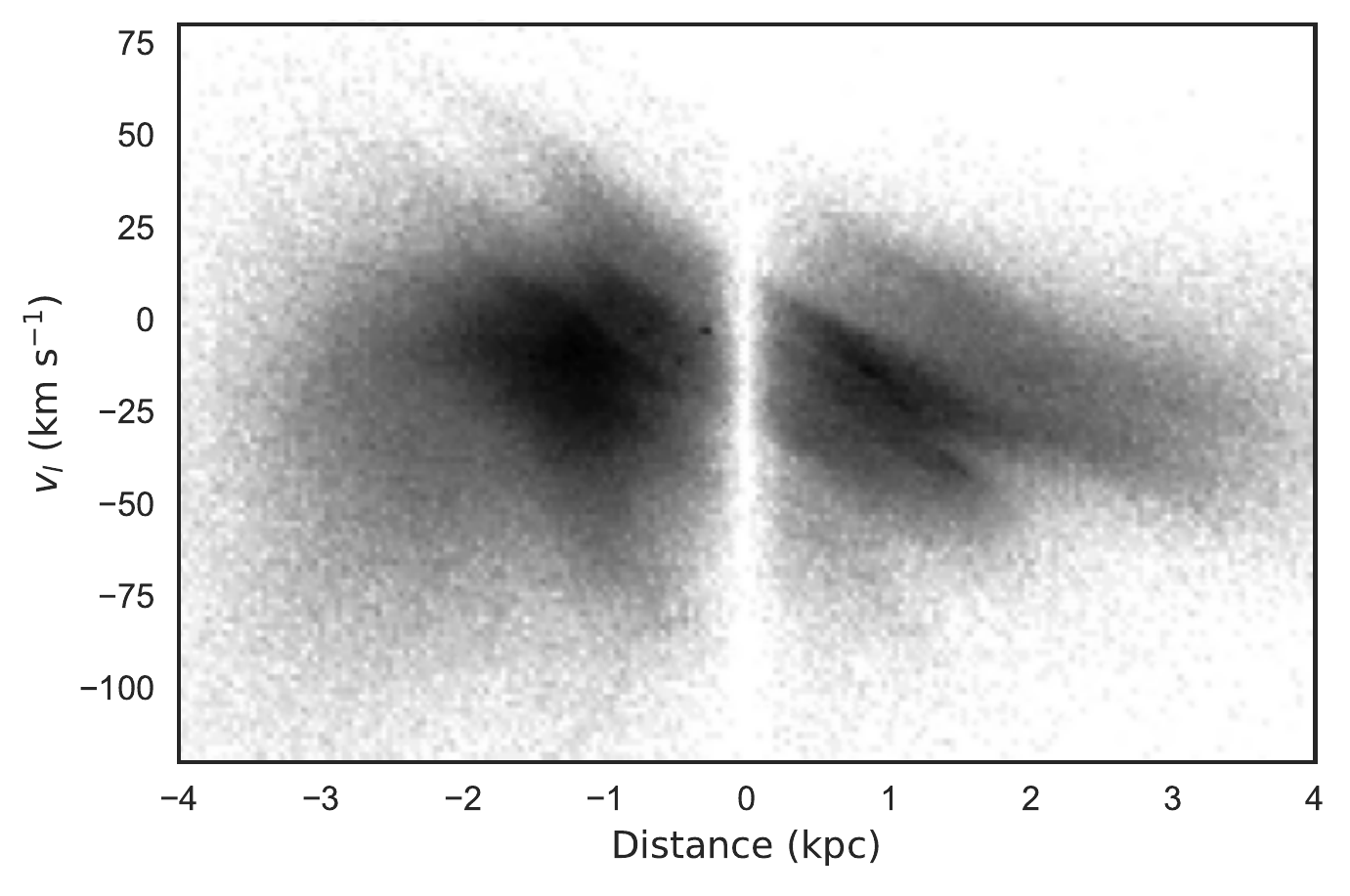}
\caption{$Upper$: $v_{\text{R}}$-$v_{\phi}$ plane in the Solar neighbourhood. We assume $R_0=8$ kpc, $V_{\mathrm{circ}}=220$ km s$^{-1}$, $U_{\odot}=-10$ km s$^{-1}$ and $V_{\odot}=24$ km s$^{-1}$. $Lower$: Distribution of $v_l$ (km s$^{-1}$) as a function of distance from the Sun. The distance is negative in the direction of the Galactic centre}
\label{RvVPhi}
\end{figure}

We then weight the amplitude with a Gaussian using the \texttt{GaussianAmplitudeWrapperPotential} to control the strength of the transient arm, where the amplitude gets multiplied with the function
\begin{equation}
A(t) = \exp\left(-\frac{[t-t_0]^2}{2\,\sigma^2}\right),
\end{equation}
and $\sigma$ is the standard deviation of the Gaussian, which controls the lifetime of the transient spiral potential. Although the wings of the gaussian technically stretch to infinity, the density enhancements lasts approximately $\mathcal{L}\approx5.6\times\sigma$ from formation to disruption. This is a simple model potential to approximate the winding material arms observed in $N$-body simulations.

In \texttt{galpy} the transient corotating spiral model can be set up as, e.g., 
\begin{minted}{python}
from galpy.potential import SpiralArmsPotential, \
  CorotatingRotationWrapperPotential, \
  GaussianAmplitudeWrapperPotential
to=0.
csp= GaussianAmplitudeWrapperPotential(\
      pot=CorotatingRotationWrapperPotential(\
        pot=SpiralArmsPotential(),
        vpo=1.,to=to),
      to=to,sigma=1.) 
\end{minted}
for the \texttt{SpiralArmsPotential} with default parameters. Set up like this, the pattern looks exactly like the input \texttt{SpiralArmsPotential} at time \texttt{to} at which it also has its peak amplitude. Before and after \texttt{to} the pattern is winding up.

Figure \ref{threespirals} shows an example of the spiral density enhancement for a model created with the \texttt{SpiralArmsPotential} within a \texttt{CorotatingRotationWrapperPotential}, weighted by the \texttt{GaussianAmplitudeWrapperPotential}. The three panels show the same arm in it's growth phase (upper), the midpoint of its lifetime (centre) and in its disruption phase (lower). For our spiral model we set $N=2$, $R_s=0.3$, $C_n=1$, $H=0.125$ and the pitch angle at the present as $\theta_{\mathrm{sp}}=12$ deg, to roughly correspond to the average measurement of the pitch angle of the Perseus arm \citep{Vallee15}. The reference angle is set such that the Solar position is approximately 2 kpc interior of the Perseus arm at $l=180$ at $t=0$ (see centre panel of Figure \ref{threespirals}). 

We set the peak amplitude of the spiral to be $\pm0.0136$ $M_{\odot}$ pc$^{-3}$ which corresponds to a relative density contrast of 1.31 between the arm and interarm region when considering a disc density of 0.1 $M_{\odot}$ pc$^{-3}$ at the Solar radius, taken from \sc{MWPotential2014 }\rm potential in \sc{galpy }\rm which in turn was fit to the measurement of $0.1\pm0.01$ $M_{\odot}$ pc$^{-3}$ from \cite{HF00}. This is similar to the values found by \cite{DS01} who find a contrast of 1.32, or \cite{Betal05} who find a contrast of 1.3. However, note that the contrast of 1.31 in the model occurs at the peak of the gaussian which controls the amplitude of the spiral. However, it is unlikely that we are observing the Milky Way spiral structure at its peak. E.g. \cite{BKMGH18} and \cite{TPZ18} find the Perseus arm is currently in the disruption phase, which implies the contrast would be stronger in the past. Thus, it is likely that we are underestimating the strength of the spiral perturbation. \cite{GPP04} find a range of 1.2-1.6 for a sample of external spiral galaxies, so the model falls well within the observed range of possible amplitudes regardless of the current state of the Milky Way spiral structure. The method of calculating the arm-interarm density contrast for the different studies is summarized in \cite{Antoja+11}.

As an initial demonstration of the effect of a winding transient arm on the kinematics of the Solar neighbourhood, we use the test particle backward integration technique detailed in \cite{D00} to construct models where a single winding arm is allowed to form and disrupt, with a range of lifetimes, $\mathcal{L}$, and formation times, $t_{\mathrm{form}}$,  with respect to the present. Figure \ref{LvsF} shows 36 models with varying lifetimes (40 to 240 Myr, top to bottom) and varying formation times, shown at the peak of the density of the gaussian (-1.72 to -0.29 Gyr in the past, left to right). The formation time, $t_{\mathrm{form}}=t_{\mathrm{peak}}-\mathcal{L}/2$. E.g. the top left model formed the longest ago, and is the shortest lived, and the lower right model formed the most recently and, and is the longest lived. The arms are all disrupted by the current time, $t=0$, yet their effect on the velocity distribution lives on for at least 1 Gyr after their amplitude peaks. E.g. the effects on the velocity distribution for these models are caused by phase wrapping \citep{MQWFNSB09} after the perturbation from the spiral arm, as opposed to the distinct resonances which would be occur for a density wave arm or mode with a fixed pattern speed.

Even for a single iteration of a transient spiral arm, the range of possible impacts on the Solar neighborhood kinematics is large. For example, the models in the second row (panels 7-12), with a lifetime of $\sim80$ Myr, show many of the arch features identified in e.g. \cite{Antoja+18}. Many of the models across a range of ages and lifetimes provide Hercules like features when combining with the CR of the long bar, some with a double density peak which is seen in $Gaia$ DR2 \citep[e.g.][]{Antoja+18,TCR18}. In addition, the longer lived arms, in the lower rows, reproduce the tilt in the velocity distribution.

As mentioned above, the structure in the $v_{\text{R}}$-$v_{\phi}$ plane arises from phase wrapping after the spiral perturbation. Figure \ref{presentday} shows the $v_{\text{R}}$-$v_{\phi}$ plane for a spiral which peaks at the present day, $t=0$, matching the centre panel of Figure \ref{threespirals}, with lifetime 40 Myr (left panel) and 240 Myr (right panel). Both spiral arms have very little effect on the $v_{\text{R}}$-$v_{\phi}$ plane, although the longer lived arm does cause the tilt in the distribution.

\begin{figure*}
\centering
\includegraphics[width=\hsize]{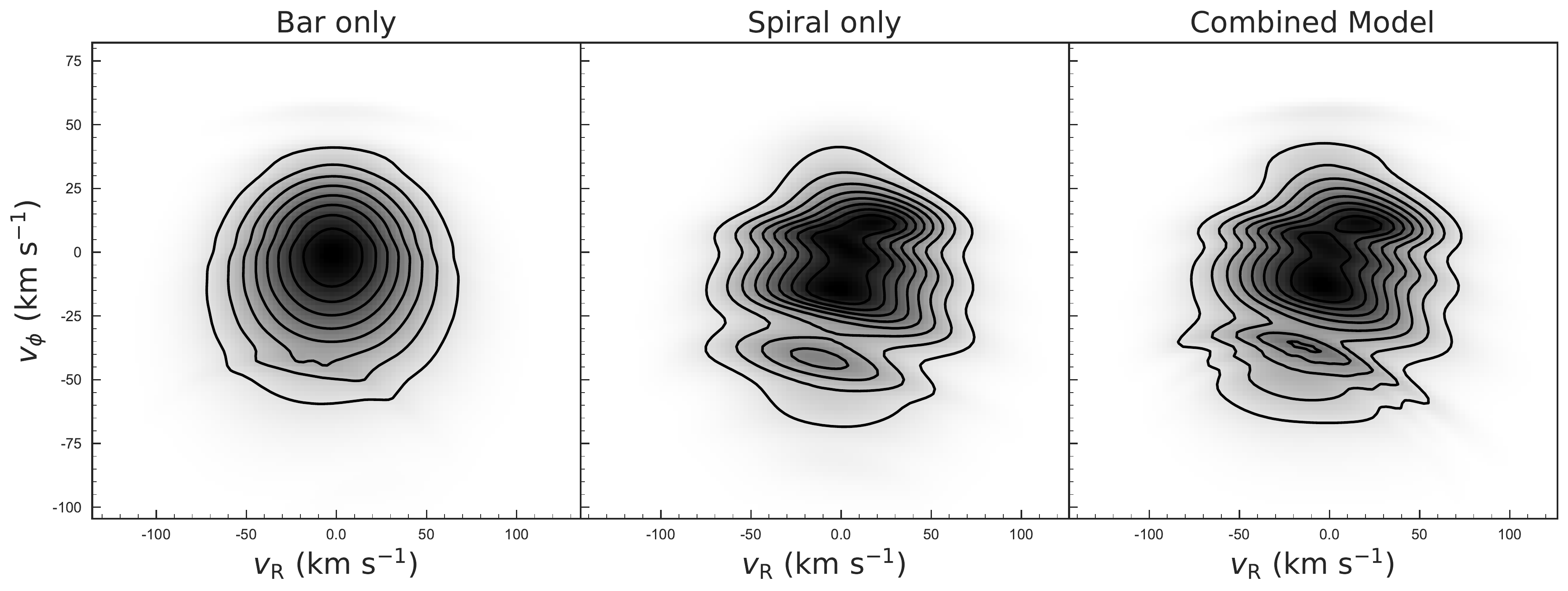}
\caption{$v_{\text{R}}$-$v_{\phi}$ plane in the Solar neighbourhood for the bar model alone (left), the spiral model alone (centre), and the combined model (right)}
\label{combi-corot}
\end{figure*}

However, as seen commonly in $N$-body simulations, the transient winding arms do not only form and wind up once, but are continually reforming features within the disc. We do not expect any of the panels in Figure \ref{LvsF} to fully reproduce the velocity distribution observed in the $Gaia$ data, because the present day kinematic structure will bear the impression of multiple spiral arms which formed and disrupted hundreds of millions of years in the past. Thus, we combine a series of recurring transient spirals to model the kinematics of the Solar neighbourhood.

\section{Comparison with the Solar neighborhood velocity distribution}
\label{thedata}
\subsection{The data}

In this section, we show the velocity distribution observed in $Gaia$ DR2, previously explored by various authors \cite[e.g.][]{GCKatz+18,KBCCGHS18,Antoja+18,RAF18,TCR18}, and identify the features we later recover.

The upper panel of Figure \ref{RvVPhi} shows the distribution of radial and azimuthal velocities for stars within 200 pc, and fractional parallax error of less than 10 per cent. We naively calculate the distance $d=1/\pi$ which is relatively safe only at low fractional parallax error. For the calculation of $v_{\text{R}},\ v_{\phi}$, we assume $R_0=8$ kpc, $v_{\text{circ}}=220$ km s$^{-1}$, $U_{\odot}=-10$ km s$^{-1}$ \citep{Betal12} and $V_{\odot}=24$ km s$^{-1}$ \citep{2015ApJ...800...83B}. This is the \emph{Gaia} DR2 view of the much studied local velocity distribution. 

The lower panel of Figure \ref{RvVPhi} shows the distribution of $v_{l}=\mu_l\times4.74047/\pi$ (km s$^{-1}$, as a proxy for $v_{\phi}$ as shown previously in \cite{Hunt+16} and \cite{KBCCGHS18}) as a function of distance from the Sun along the Galactic Centre-Sun-Galactic anticentre line. Here we select stars brighter than $G<15.2$ mag, with $-10\leq b\leq10$ deg and either $-10\leq l\leq 10$ deg or $170\leq l\leq 190$ deg. We then select stars with fractional parallax error of less than 15 per cent, $\mid z\mid\leq0.5$ kpc and with $v_b=\mu_b\times4.74047/\pi\leq20$ km s$^{-1}$.   

The `ridges' identified in \cite{KBCCGHS18}, \cite{Antoja+18}, \cite{RAF18} and \cite{Quillen+18} are clearly visible. The exact choice of sampled area, quality cuts and method has some effect on how clearly visible the features are, but they remain consistent between the works.

If the ridges are caused by radial migration at specific resonances, e.g. the bar and spiral CR or OLR, then there will be a limited number of ridges, each associated with a resonance. This is illustrated nicely in Figure 4 of \cite{Antoja+18}. However, we observe more ridge features than are easily explained by the combination of resonances arising from a potential component with a fixed pattern speed. In addition, as noted in \cite{RAF18}, while some of the ridges conserve their vertical angular momenta, which would be expected for stars on resonant orbits, e.g. Hercules, some do not e.g. Sirius. 

The horizontal phase mixing shown in \cite{Antoja+18} could account for the features with a non-resonant origin, and also creates significantly more, yet weaker, ridges which better represent the data. This assumes the Milky Way is still phase mixing after an event such as the perturbation of the disc by the recent passage of the Sagittarius dwarf galaxy. However, the phase wrapping from their model does not account for the tilt in the $v_{\text{R}}$-$v_{\phi}$ plane \citep[as also noted in][]{Quillen+18}, which appears to be well reproduced by the transient-spiral arms model shown below.

\begin{figure*}
\centering
\includegraphics[width=\hsize]{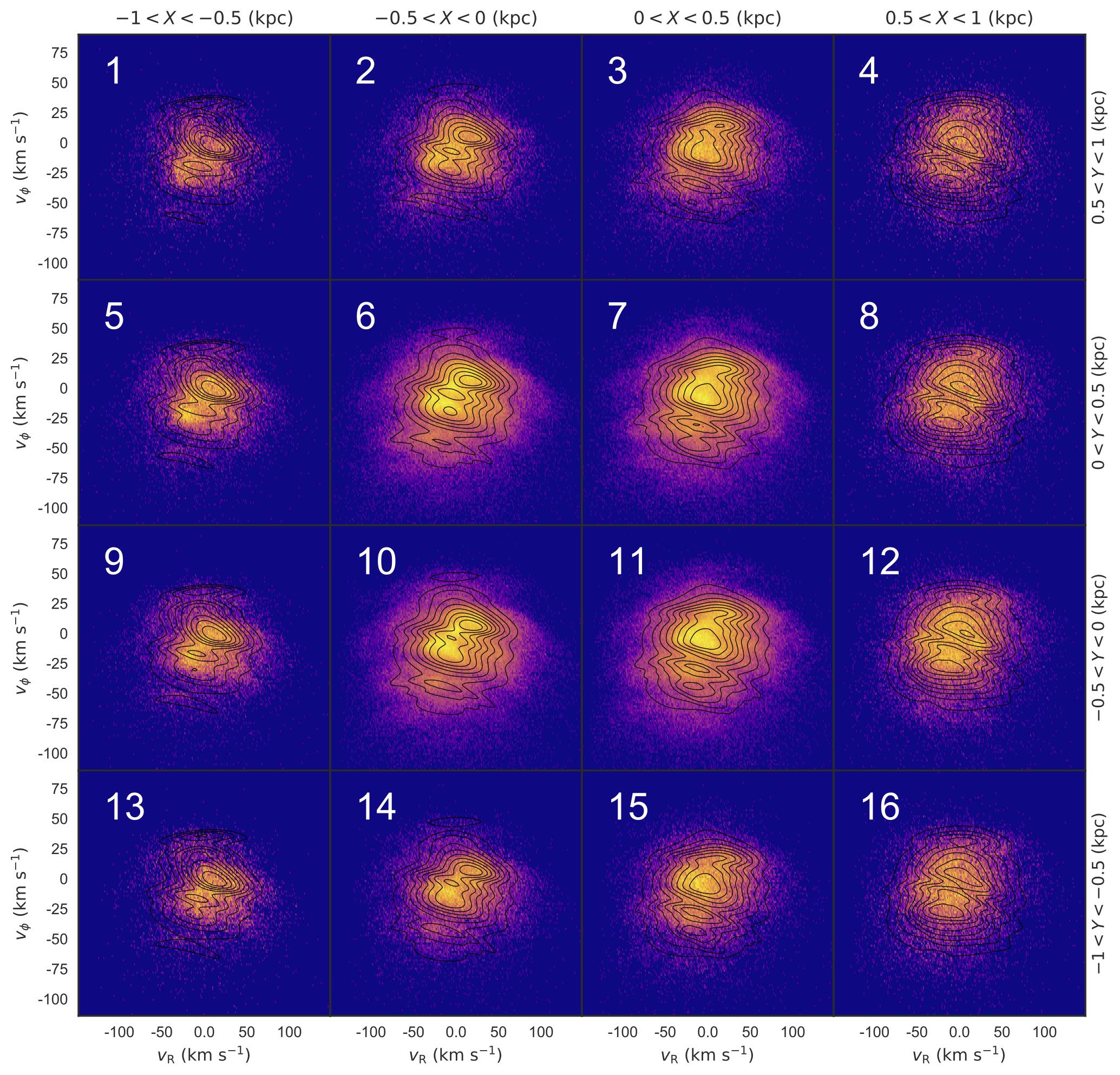}
\caption{$v_{\text{R}}$-$v_{\phi}$ planes in 500 pc bins nearby the Solar neighbourhood, for $-1\leq X\leq-0.5$ (left column), $-0.5\leq X\leq0$ (second column), $0\leq X\leq0.5$ (third column), $0.5\leq X\leq1$ (right column), $0.5\leq Y\leq1$ (top row), $0\leq Y\leq0.5$ (second row), $-0.5\leq Y\leq0$ (third row) and $-1\leq Y\leq-0.5$ (bottom row) for the model (contours) and the data (color map density plot). The numbers exist only to allow individual panels to be referenced in the analysis.}
\label{16UVs}
\end{figure*}

\subsection{The models}
In the model, the spiral structure is set to corotate with the stars. We combine a long slow bar potential, with the corotating winding potential described in Section \ref{spiral}. Although we use the long bar model here, the short bar is easily able to reproduce the main bifurcation in the $v_{\text{R}}$-$v_{\phi}$ plane corresponding to Hercules as shown in numerous other works \citep{D00}. We choose the long bar here to demonstrate how easily the combination of spiral structure with the long bar potential also reproduces the Hercules stream.

To mimic the transient reforming nature of the corotating spiral arm, we set a series of three corotating spirals, with a lifetime of $\sim250$ Myr, which occur $\sim225$ Myr apart, with the amplitude of the first spiral peaking $\sim450$ Myr in the past, and the third one peaking today at $t=0$ in the location of the Perseus arm. Note that this is a very regular series of arm formation, whereas $N$-body models show more diverse structure in terms of arm strength, and formation time \citep[e.g.][]{GKC13}. 

Figure \ref{combi-corot} shows the $v_{\text{R}}$-$v_{\phi}$ plane in the Solar neighbourhood for the bar model alone (left), the spiral model alone (centre), and the combined model (right). The left panel shows only a minor perturbation in the area of the Hercules stream. This originates from the corotation resonance of the long slow bar, as explored in other works \citep[e.g.][]{P-VPWG17}. The centre panel shows the $v_{\text{R}}$-$v_{\phi}$ plane for a model with only the winding spiral structure, which surprisingly reproduces Hercules nicely, without any influence from the bar. The right panel shows the model where the bar and spiral potentials are combined. The interaction of the bar and spiral potentials slightly modifies the shape of the velocity distribution, but in this model, the majority of the features come from the spiral arm potential. This is similar to what was found in \cite{QDBMC11}, who showed that transient spiral waves reproduce the tilt in the velocity distribution, and can lead to bifurcations similar to Hercules in the outer disc.

The model shows a decent recovery of the kinematics in the Solar neighbourhood, with a clear Hercules like feature, arising from the corotating spiral structure, and also retains the striated features in the main model of the velocity distribution roughly corresponding to the other moving groups. Note that we do not expect a perfect recovery of the $v_{\text{R}}$-$v_{\phi}$ plane, because the exact combination of the number, frequency, lifetime, pitch angle and strength of the transient spiral arms will make significant differences. Our aim here is to show that we are able to reproduce the kinematics in a qualitative sense via the repeated perturbation from winding arms. We also demonstrate that it is possible to reproduce Hercules without any bar influence, which is an interesting result in itself. However, as the Milky Way is known to be barred, we use the bar+spiral potential for the subsequent analysis.

With the data from \emph{Gaia} DR2, we can compare the $v_{\text{R}}$-$v_{\phi}$ plane not only in the Solar neighbourhood, but slightly further across the disc \citep[e.g. as shown in][]{GCKatz+18}. Figure \ref{16UVs} shows the model (contours) overlaid on the data (color map density plot) for the $v_{\text{R}}$-$v_{\phi}$ plane at 16 different 500 by 500 pc bins. They are (in kpc): $-1\leq X\leq-0.5$ (left column), $-0.5\leq X\leq0$ (second column), $0\leq X\leq0.5$ (third column), $0.5\leq X\leq1$ (right column), $0.5\leq Y\leq1$ (top row), $0\leq Y\leq0.5$ (second row), $-0.5\leq Y\leq0$ (third row) and $-1\leq Y\leq-0.5$ (bottom row). The contours, and the color map both track the stellar density.

Figure \ref{16UVs} shows that the model contours reproduce the tilt of the velocity distribution, and the presence of multiple moving groups. Panels 6 and 10 show clearly three moving groups in the main mode of the distribution (e.g. `above' the Hercules bifurcation), which are qualitatively similar to Pleiades, Coma Berenices and Sirius, along with a distinct Hercules. Panels 7 and 11 show that a secondary mode within Hercules forms once we move slightly inwards from the Solar radius \citep[as observed in][]{Antoja+18,RAF18,TCR18} and also matches the shift in the main bifurcation to higher $v_{\phi}$ (km s$^{-1}$). The outer ring of panels contain less stars, and it is harder to see the trend in the change of moving groups, other than the main bifurcation between Hercules and the main mode, which is well traced in $v_{\phi}$ across all panels. However, the tilt in the bifurcation is slightly too shallow in the models in the right column, e.g. towards the Galactic centre, when compared with the data. 

In addition to examining the $v_{\text{R}}$-$v_{\phi}$ plane, we also examine the distribution of rotation velocities as a function of radius, and test of recovery of the `ridges' in this projection. We calculate the distribution of $v_{\phi}$ every 100 pc in the Galactocentric radius $4-12$ kpc along the GC-Sun-GA line-of-sight.

\begin{figure}
\centering
\includegraphics[width=\hsize]{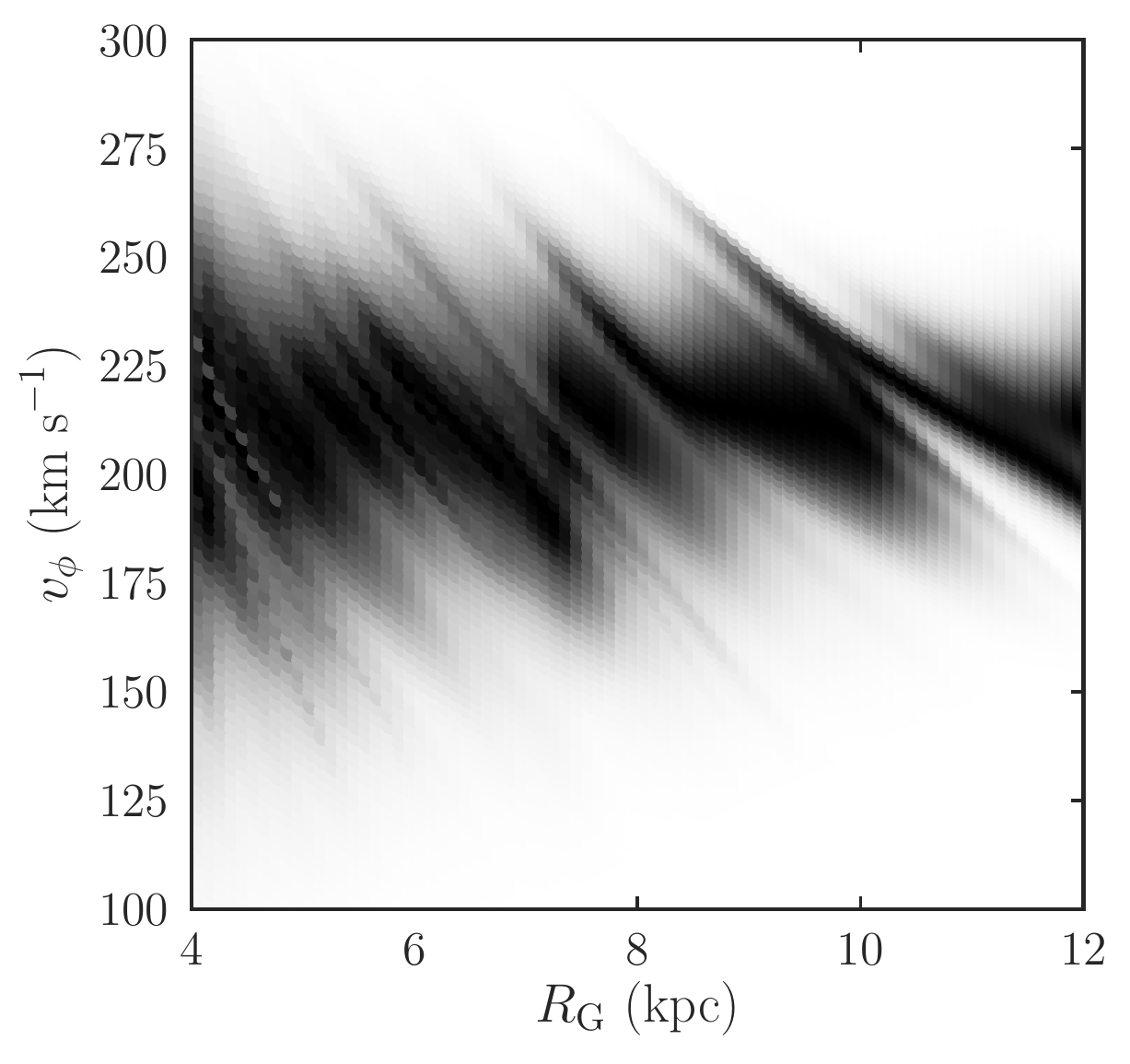}
\caption{Distribution of $v_{\phi}$ as a function of $R$ in the model. The ridge features observed in $Gaia$ DR2 are clearly visible in the model.}
\label{tpmodels}
\end{figure}

Figure \ref{tpmodels} shows the distribution of rotation velocities as a function of Galactic radius along the GC-Sun-GA line for our model. The presence of multiple ridges is clear, and the angle of tilt is similar to what is seen in the data, e.g. the velocity of the ridges decreases around 25-30 km s$^{-1}$kpc$^{-1}$ around the Solar radius.

We can see the double Hercules-like feature around 8 kpc from the Galactic centre. The large split around 11 kpc is the OLR resonance feature from the long bar model used here. Note that we do not observe any such large split in the data around 11 kpc. However, the quantity and quality of the data drops quickly with distance from the Sun, 

We are not suggesting that the spiral arm model presented here is the correct parameterisation of the Milky Way's spiral arms, merely that a series of corotating transient spiral arms naturally leads to the ridge features seen in the $Gaia$ DR2 data. However, there are other potential explanations for the ridge and arch structure as mentioned above, e.g. phase wrapping after the close passage or merger of a dwarf galaxy \citep[e.g.][]{MQWFNSB09} or the combination of many individual resonances \citep[e.g.][]{Antoja+18}, or the recent crossing of spiral arms \citep{Quillen+18}.

While it may be possible to fit the individual ridges to specific occurrences of spiral structure, similar to what is done in \cite{Quillen+18}, we defer this to a future work, as the parameter space to be explored is large. E.g., we are not only constrained to fit the features to spiral arms which are visible in the Galaxy today, but also those which were disrupted in the recent past. This is thus worth noting that we should be careful when trying to reproduce all the kinematic features in the $Gaia$ data with current structure. If the Milky Way's spiral arms are transient, winding and recurrent, there is likely no direct link between some of the kinematic substructure and present day spiral arms.

In addition to reproducing the ridges, we note that we are also able to explain the Hercules stream as a result of the transient spiral structure, either alone, or in combination with the corotation resonance of the long slow bar. \cite{Hattori+18} showed a similar result for the density wave spiral model. These new developments make it difficult to determine the pattern speed or length of the bar via fitting to the Hercules stream alone.

\section{Discussion and outlook}\label{summary}
In this work we show that the transient winding spiral arms commonly seen in $N$-body simulations naturally reproduce the ridges and arches observed by $Gaia$ DR2 in the stellar kinematics, without needing to invoke perturbation of the disc via an external force such as the recent passage of the Sagittarius dwarf galaxy.

We also show that it is relatively straightforward to create a model of the Hercules stream from the winding spiral arm potential, either with or without the presence of a bar. Our model for the Milky Way transient spiral potential creates a distinct feature in the kinematic area of Hercules. However, note that this is not explicit evidence against the classical short fast bar, which also reproduces Hercules well without including spiral structure.

Further work is needed to try and fit individual ridges observed in the $Gaia$ data to specific occurrences of the transient spiral structure, such as is done in \cite{Quillen+18} for the model involving overlapping density waves. If the Milky Way's spiral structure is indeed transient and winding, it may allow us not only to reproduce the current spiral structure, but also provide a window on the previous generations of spiral arms.

\section*{Acknowledgements} We thank A. Quillen for helpful comments which improved the manuscript. J. Hunt is supported by a Dunlap Fellowship at the Dunlap Institute for Astronomy \& Astrophysics, funded through an endowment established by the Dunlap family and the University of Toronto. J. Hong and J. Bovy received support from the Natural Sciences and
Engineering Research Council of Canada (NSERC; funding reference number RGPIN-2015-05235). J. Bovy also received partial support from an Alfred P. Sloan Fellowship.
D. Kawata acknowledges the support of the UK's Science \& Technology Facilities Council (STFC Grant ST/N000811/1). RJJG acknowledges support by the DFG Research Centre SFB-881 `The Milky Way System', through project A1. This project was developed in part at the 2018 NYC $Gaia$ Sprint, hosted by the Center for Computational Astrophysics of the Flatiron Institute in New York City. This research made use of Astropy, a community-developed core Python package for Astronomy \citep{Astropy}. This work has also made use of data from the European Space Agency (ESA) mission Gaia (https://www.cosmos.esa.int/gaia), processed by the Gaia Data Processing and Analysis Consortium (DPAC, https://www.cosmos.esa.int/web/gaia/dpac/consortium). Funding for the DPAC has been provided by national institutions, in particular the institutions participating in the Gaia Multilateral Agreement.

\bibliographystyle{mn2e}
\bibliography{ref2}

\label{lastpage}
\end{document}